\documentstyle[prd,floats,aps]{revtex}
\begin{document}
\title{Nonexistence of  conformally flat slices of the Kerr spacetime}
\author{Alcides Garat$^1$,
Richard H. Price$^2$\\
1. {\it Department of Physics, University of Utah, Salt Lake City, Utah
84112. On leave from Universidad de la Rep\'ublica, Montevideo, Uruguay.}\\
2. {\it Department of Physics, University of Utah, Salt Lake City, Utah
84112.}}

\maketitle
\begin{abstract}

Initial data for black hole collisions are commonly generated using
the Bowen-York approach based on conformally flat 3-geometries.  The
standard (constant Boyer-Lindquist time) spatial slices of the Kerr
spacetime are not conformally flat, so that use of the Bowen-York
approach is limited in dealing with rotating holes. We investigate
here whether there exist foliations of the Kerr spacetime that are
conformally flat. We limit our considerations to foliations that are
axisymmetric and that smoothly reduce in the Schwarzschild limit to
slices of constant Schwarzschild time. With these restrictions, we
show that no conformally flat slices can exist.

\end{abstract}
\vspace{-8.5cm} 
\begin{flushright}
\baselineskip=15pt
CGPG-97/12-?  \\
gr-qc/???\\
\end{flushright}
\vspace{7.5cm}

\section{Introduction}

Perhaps the most exciting source that might be detected by
gravitational wave detectors now in development is radiation from a
merger of black holes. This has been one of the motivations for the
effort being put into the application of numerical relativity to black
hole collisions. In this work supercomputers are used to evolve
initial value solutions of Einstein's field equations. The computation
of the initial value solutions is itself a difficult task, and much of
the work has taken advantage of the Bowen-York \cite{BowenYork}
program for initial value solutions, in which the restriction is made
that the initial 3-geometry is conformally flat.

Despite the elegance and convenience of the Bowen-York approach, a
conformally flat initial solution has a serious shortcoming for work
with black holes.  Astrophysically realistic black holes will be
rapidly rotating. The spatial geometry of the Kerr spacetime of
rotating holes is not conformally flat. More specifically, for Kerr
spacetime described in standard Boyer-Lindquist\cite{boyerlindquist}
coordinates $t,r,\theta,\phi$, a slice at constant $t$ is not
conformally flat.  Because of this, the use of Bowen-York initial data
to study colliding holes entails two difficulties. First, the
Bowen-York representation of a rotating hole will be that of a
distorted Kerr hole. Numerical evolution of this solution will produce
a burst of radiation as each of the colliding holes ``relaxes'' to an
approximately Kerr form\cite{singlespin}. If it were possible to start
the collision of the holes at large separation, this initial burst
would be easily distinguishable from the radiation arising from the
merger itself. But for the present, numerical evolutions must start quite
close to the final state of the merger.  The second difficulty is that
the Bowen-York program cannot give an initial value solution that is a
perturbation of the final single stationary hole, since that final
solution does not have conformally flat spatial slices. This precludes
the application of ``close limit'' perturbation theory that has proven
to be very successful as an approximation scheme for black hole
mergers\cite{closelimit}.

It would be extremely useful if the need to deal with Kerr black holes
could be reconciled with the convenience of the Bowen-York conformally
flat scheme. For that reason, we inquire here whether there might
exist conformally flat slices of the Kerr spacetime. We know that the
$t=$\,constant slices are not flat, but it may be that the geometry of
a different kind of slicing, of the form $t=f(r,\theta,\phi)$ {\em
is} conformally flat. The question of conformal flatness of any 3-geometry
can most conveniently be answered with the use of the 
the Bach (Cotton-York) tensor\cite{KSMH},
\begin{eqnarray}\label{Bachdef} 
B^{ij}&=& 2 \epsilon^{ikp}
[R^{j}_{k}-\frac{1}{4}\ \delta^{j}_{k}R]_{;p}\ ,
\end{eqnarray}
where $R^{j}_{k}$ is the Ricci tensor for the 3-geometry, $R\equiv
R^{k}_{k}$ is its Ricci scalar, and the semicolon refers to covariant
differentiation with respect to the metric of the 3-geometry. The
vanishing of the Bach tensor is a necessary and sufficient condition
for a 3-geometry to be conformally flat.

Using the Bach tensor, we investigate here whether a conformally flat
slice of the Kerr spacetime might exist, but we make certain
restrictions on the search that limit the generality of our
conclusion. One restriction is that we consider only axisymmetric
slices of the axisymmetric Kerr geometry. One reason for this is
practical: Gaining the advantages of conformal flatness while losing
axisymmetry would be a Pyrrhic victory.  A second reason is that the
extension of our conclusion to a nonaxisymmetric slicing turns out to
be quite difficult. We restrict ourselves, therefore, to slices of the
form $t=f(a,r,\theta)$, where $a$ is the Kerr spin parameter.

A second restriction we make is to consider only families of slicings
that have the property that in the Schwarzschild ($a\rightarrow0$)
limit, the slicings smoothly go to slices of constant 
Schwarzschild time. This means that the slicing function
$f(a,r,\theta)$, would have the limit zero as $a\rightarrow0$, since
the Boyer-Lindquist coordinates become the Schwarzschild coordinates
as ($a\rightarrow0$). We assume, furthermore, that $f$ can be expanded
in $a$ so that
\begin{equation}\label{sliceq} 
t=aF(r,\theta)+{\cal O}(a^2)\ .
\end{equation}
Our approach is to compute the Bach tensor for the 3-geometry induced
by the slicing in eq. (\ref{sliceq}) and to expand the Bach tensor in $a$.
We will show that no slicing can be found that makes the tensor vanish
to lowest nontrivial order in $a$, and we conclude that no family of
slices of this type can be conformally flat.

The assumption in eq.\,(\ref{sliceq}) means that our conclusion does
not rule out  a conformally flat slicing
for an isolated value of $a$, or  a family of conformally
flat slicings for a range of $a$ that does not include $a=0$. These
limitations are inherent to our method of expanding about $a=0$. A more
subtle shortcoming of our method is that it does not rule out a family
of slicings of the form
\begin{equation}\label{gensliceq} 
t=G(r,\theta)+aF(r,\theta)+{\cal O}(a^2)\ .
\end{equation}
where $G$ is not a constant.
That is, it does not rule out a family of slicings that, in the
Schwarzschild ($a\rightarrow0$) limit takes the form $t=G(r,\theta)$.
But there {\em are} slicings of Schwarzschild other than
$t=$\,constant that are conformally flat. In fact any spherically
symmetric 3 geometry is conformally flat, so any slicing of the form
$t=G(r)$ is conformally flat. Such slicings of Schwarzschild, of
course, are not orthogonal to the timelike Killing vector for the
spacetime geometry. The use of such a slicing would have disadvantages
for initial data similar to the disadvantage of a nonaxisymmetric
slicing.  For the Kerr spacetime of course, the Killing vector is not
hypersurface orthogonal, so there is no foliation that is singled out
by the symmetry. Still, intuition suggests that a useful slicing for
Kerr should have extrinsic curvature that only describes the rotation, and
that vanishes as $a\rightarrow0$. This is equivalent to a slicing that 
reduces to one of constant Schwarzschild time.

\section{Kerr metric tensor}
In Boyer-Lindquist\cite{boyerlindquist} coordinates, the components
of the Kerr metric have the explicit form
\begin{eqnarray}
^{(4)}g_{tt} &=& -1+2Mr/(r^2+a^2\cos^2\theta) \label{METRICstart}\\
^{(4)}g_{t\phi} &=& -2Mra\sin^2\theta/(r^2+a^2\cos^2\theta)\\
^{(4)}g_{rr}&=& (r^2+a^2\cos^2\theta)/(r^2-2Mr+a^2)\label{grr}\\
^{(4)}g_{\theta\theta} &=&  r^2+a^2\cos^2\theta\label{METRICpenult}\\
^{(4)}g_{\phi\phi} &=&
[(r^2+a^2)^2-(r^2-2Mr+a^2)a^2\sin^2\theta]\sin^2\theta
/(r^2+a^2\cos^2\theta)\label{METRICend}\ .
\end{eqnarray}
The 3-geometry $^{(3)}g_{ij}$ induced on a spatial slice given by
eq.\,(\ref{sliceq}) has components
\begin{eqnarray}
^{(3)}g_{rr}&=&^{(4)}g_{rr}+\,^{(4)}g_{tt}a^2\left(\partial_rF\right)^2
+{\cal O}(a^3)\label{3g1} \\ 
%%%%%%%%%%
^{(3)}g_{\theta\theta} &=&
^{(4)}g_{\theta\theta}+\,^{(4)}g_{tt}a^2\left(\partial_\theta F\right)^2
+{\cal O}(a^3) \label{3g2}\\ 
%%%%%%%%%%%%%%%%%%%%%
^{(3)}g_{r\phi}&=&2\,^{(4)}g_{t\phi}\,a\left(\partial_rF\right)
+{\cal O}(a^3)\label{3g3}\\ 
%%%%%%%%%%%%%%%%%%%%%
^{(3)}g_{\theta\phi}&=&2\,^{(4)}g_{t\phi}\,a\left(\partial_{\theta}F\right)
+{\cal O}(a^3)\label{3g4}\\ 
%%%%%%%%%%%%%%%%%%%%%
^{(3)}g_{\phi\phi} &=&^{(4)}g_{\phi\phi}\label{3g5}\ .
%%%%%%%%%%%%%%%%%%%%%
\end{eqnarray}
From these expressions, and from the fact that $g_{t\phi}$ is
proportional to $a$, it follows that the metric functions 
$^{(3)}g_{ij}$
have no
terms first order in $a$.  The deviations of 
$^{(3)}g_{ij}$ from a $t=$\,constant
slice of the Schwarzschild spacetime are therefore of order $a^2$
and higher. Since the $t=$\,constant Schwarzschild slice is 
conformally flat, and hence has a vanishing Bach tensor, the components
of the Bach tensor for eqs.\,(\ref{3g1}) -- (\ref{3g5}) vanish to first 
order in $a$.

\section{Diagonal components of the Bach tensor}

To second order in $a$, the diagonal components of the Bach tensor
turn out to be given by 
\begin{eqnarray}
B^{rr} &=& 6 M a^2(r-2M)  \left(r^5 \sqrt{r^5 \sin^2\theta /(r-2M)}\right)^{-1}
\nonumber\\&& \times\left[ 3 (\sin^2\theta-\cos^2\theta)
\ \partial_{\theta}F(r,\theta)- 5 \cos\theta \sin\theta  
\ \partial^2_{\theta}F(r,\theta) - \sin^2\theta
\ \partial^3_{\theta}F(r,\theta) \right]\label{Brr}\\
B^{\theta\theta} &=&6 M a^2 \sqrt{r^5  /(r-2M)} 
\left(r^{12} \sin\theta\right)^{-1} \nonumber\\&&
\times\left[ (-r^4+4M^2r^2\cos^2\theta+r^4\cos^2\theta
-4Mr^3\cos^2\theta-4M^2r^2+4Mr^3)\ \partial_{\theta}\partial^2_{r}F(r,\theta) + 
\right. \nonumber\\&&
(-8r^2-56M^2+11r^2\cos^2\theta-50Mr\cos^2\theta+56M^2\cos^2\theta+44Mr)
\ \partial_{\theta}F(r,\theta) +
\nonumber\\&& (5r^3-26M^2r\cos^2\theta+23Mr^2\cos^2\theta+
26M^2r-5r^3\cos^2\theta-23Mr^2)
\ \partial_{\theta}\partial_{r}F(r,\theta) + \nonumber\\&& \left. 
r(r-2M)\cos\theta\sin\theta
\ \partial^2_{\theta}F(r,\theta) \right] \label{Bthth}\\
B^{\phi\phi} &=& 6 M a^2 
\left(r^7 \sin^{2}\theta \sqrt{r^5 /(r-2M)}\right)^{-1} \nonumber\\&&
\times\left[ (5r-28M)\sin\theta\ \partial_{\theta}F(r,\theta) + 
4r\cos\theta\ \partial^2_{\theta}F(r,\theta) + (13M-5r)\ r\sin\theta
\ \partial_{\theta}\partial_{r}F(r,\theta) + \right. \nonumber\\&& \left. r^2(r-2M)\sin\theta 
\ \partial_{\theta}\partial^2_{r}F(r,\theta) + r\sin\theta
\ \partial^3_{\theta}F(r,\theta) \right]\label{Bphiphi}
\end{eqnarray}
These three diagonal components are not independent, but are related
by the fact that the trace $B^i_i$ vanishes, as can easily be checked
(to second order in $a$) for eqs.\,(\ref{Brr})--(\ref{Bphiphi}).

If $B^{rr}$
is to vanish, we have from eq.\,(\ref{Brr})
that
\begin{equation}
0 = \partial_{\theta}\left[ 3 \: \cos\theta \: \sin\theta \:\:
\partial_{\theta}F(r,\theta) + \sin^2\theta \:\: \partial^2_{\theta}F(r,\theta)
\right]\ . 
\end{equation}
This equation can be solved by three integrations with respect to $\theta$
to give the general solution
\begin{eqnarray}
F(r,\theta) &=& {h(r) \over 2\sin^2\theta} + u(r) 
\left[ {-\cos\theta \over 2\sin^2\theta} + \frac{1}{2}
\ln(\tan\left[\theta/2\right]) \right] + v(r)\label{F}\,,
\end{eqnarray}
in which $h(r)$, $u(r)$ and $v(r)$ are the  ``constants'' 
introduced in the three integration steps.
When this form of $F(r,\theta)$ is put into the right hand
side of eq.\,(\ref{Bthth}), the equation $B^{\theta\theta}=0$ takes
the form
\begin{equation}
0 = \hat{O}_{u}u(r) + \cos{\theta} \:\: \hat{O}_{h}h(r)\ ,
\end{equation}
and the vanishing of $\hat{O}_{h}h(r)$ and of $\hat{O}_{u}u(r)$ yield the two
differential equations
\begin{eqnarray}
0 &=& (9r^2+56M^2-46Mr) \: h(r) + (-26M^2r+23Mr^2-5r^3) \: \partial_{r}h(r) 
+ (-4Mr^3+r^4+4M^2r^2) \:  \partial^2_{r}h(r) \label{heq}\\
0 &=& (-8r^2+44Mr-56M^2) \: u(r) + (26M^2r+5r^3-23Mr^2) \: \partial_{r}u(r) 
+ (-4M^2r^2-r^4+4Mr^3) \:  \partial^2_{r}u(r) \label{ueq}\ .
\end{eqnarray}
Finally, the solutions to these equations are
\begin{eqnarray}
h(r) &=& A\ { r^{7/2} \over \sqrt{r-2M}} + B\ {
 r^{7/2}\cosh^{-1}(\sqrt{r/2M})  \over  \sqrt{r-2M}}\label{h}\\
u(r) &=& C\ { (r-M) r^{7/2} \over M \sqrt{r-2M}} + 
 D\ {r^{4} \over M }\label{u}\ .
\end{eqnarray}
At this point we have reduced the freedom in 
$F(r,\theta)$ 
to the constants $A$,$B$,$C$,$D$ and the function $v(r)$, if the
Bach tensor is to vanish.
\section{Offdiagonal components of the Bach tensor}

When the form of $F(r,\theta)$ required by eqs.\,(\ref{F}), (\ref{h}),
and (\ref{u}) is used, we find that the component $B^{r\theta}$ is
\begin{equation}
B^{r\theta}= 3a^2B\,M\,r^{-9/2}\sqrt{r-2M}/ \sin\theta \label{Erth}\ .
\end{equation}
It follows that a conformally flat slicing requires that $B=0$.

With $B$ set to zero in the form of $F(r,\theta)$ required by eqs.\,(\ref{F}), (\ref{h}),
and (\ref{u}), the rather lengthy expression for $B^{\theta\phi}$ can 
be written as
\begin{equation}
B^{\theta \phi} = a^2 \: \Xi \: \left(\sin^{15}\theta \:  r^{12}\:
(r-2M)^{3}\: M^2 
\right)^{-1}\ ,
\end{equation}
where $\Xi$ is
\begin{eqnarray}
\Xi&=& c_{1}(r) \: \cos\theta + c_{2}(r) \: \cos^2\theta + \cdots + 
c_{16}(r) \: \cos^{16}\theta \nonumber\\&&
+
\left[b_{1}(r) \: \cos\theta + b_{2}(r) \: \cos^2\theta + \cdots + 
b_{12}(r) \: \cos^{12}\theta
\right]\ln\left(\sin\theta/[1+\cos\theta]\right)
\label{INDEXP}
\end{eqnarray} 
For $B^{\theta \phi}$ to vanish, each of the terms $c_{1}(r)\cdots
c_{16}(r), b_{1}(r)\cdots b_{12}(r)$ must vanish. The sum of $b_k$
terms is explicitly
\begin{eqnarray}
\sum_{k}b_{k}(r)\cos^{k}\theta&=& (-2)\:(1-\cos^2\theta)^5
\:P_{b}(r,\theta)\nonumber\ ,
\end{eqnarray}
with
\begin{eqnarray}
P_b&=&\left[\left(30Mr^{15}-246M^{2}r^{14}+777M^{3}r^{13}
-1158M^{4}r^{12}+780M^{5}r^{11}-168M^{6}r^{10}\right)\cos^2\theta\right. \nonumber\\&&
\left.+10Mr^{15}-82M^{2}r^{14}+259M^{3}r^{13}-386M^{4}r^{12}+260M^{5}r^{11}
-56M^{6}r^{10}\right] \: DA \nonumber\\
&&- \: \left[\left(-30Mr^{29/2}+216M^{2}r^{27/2}-576M^{3}r^{25/2}
+672M^{4}r^{23/2}-288M^{5}r^{21/2}\right)\cos^2\theta\right. \nonumber\\&&
\left.-10Mr^{29/2}+72M^{2}r^{27/2}-192M^{3}r^{25/2}
+224M^{4}r^{23/2}-96M^{5}r^{21/2}\right] \: 
\sqrt{r-2M} \: CA \nonumber\\&&
+ \: [-48r^{31/2}+388Mr^{29/2}-1208M^{2}r^{27/2}+1776M^{3}r^{25/2}
-1184M^{4}r^{23/2}+256M^{5}r^{21/2}]\cos\theta \: \:
\sqrt{r-2M}  \: D^2  \nonumber\\
&&+ \: [-48r^{31/2}+388Mr^{29/2}-1224M^{2}r^{27/2}+1872M^{3}r^{25/2}
-1376M^{4}r^{23/2}+384M^{5}r^{21/2}]\cos\theta \:
\sqrt{r-2M} \: C^2  \nonumber\\&& -\: 
[96r^{16}-872Mr^{15}+3160M^{2}r^{14}-5740M^{3}r^{13}
+5320M^{4}r^{12}-2192M^{5}r^{11}+224M^{6}r^{10}]\cos\theta\: DC\ .
\end{eqnarray} 
From the terms in $P_b$ that are proportional to $\cos\theta$ it follows 
that $C$ and $D$ must vanish if $B^{\theta \phi}$ vanishes.

When the simplification $C=D=0$ is made in eq.\,(\ref{INDEXP})
the sum of the $c_{k}$ terms takes the form
\begin{eqnarray}
\sum_{k}c_{k}(r)&&\cos^{k}\theta= 84 \: M^{4} \: (r-2M)^{7/2} \:
r^{5/2}\: (1-\cos^2\theta)^8 \nonumber\\&&
- r^{7} \: M^{2}  \: (r-2M)^{3}
\: (3\cos^2\theta+1) \: (1-\cos^2\theta)^5 \nonumber\\
&&\times\left( \: [r^4-4r^3M+4r^2M^2] \: \partial_{r}^{3}v(r) 
+ [2r^3-2r^2M-4rM^2] \: \partial_{r}^{2}v(r) 
+ [-2r^2+8rM-5M^2] \: \partial_{r}v(r) \: \right)A \nonumber\\&&
+ 2 \: M^2 \: r^{19/2} 
\: (r-2M)^{7/2} \: (1-\cos^2\theta)^3 \: \nonumber\\&& 
\times[(21M-5r)\cos^4\theta-21M\cos^2\theta-3r] \: A^2\ .
\end{eqnarray}
In this expression the terms proportional to
$\cos^{11}\theta\cdots$$\cos^{16}\theta$ do not vanish for any choice of
$A$ or $v(r)$. It follows that there is no function 
$F(r,\theta)$ for which the slicing in eq.\,(\ref{sliceq}) gives a 3-geometry
that is conformally flat to second order in $a$. Under the assumptions
stated at the outset this implies that there is no slicing of the Kerr
spacetime that is conformally flat.

\section{Discussion} 

We have shown that there can be no spatial slicing of the Kerr
spacetime with the following properties: (i) The slicing is
conformally flat. (ii) The slicing is axisymmetric. (iii) The slicing,
as a function of the Kerr parameter $a$, goes smoothly to a slice of
constant Schwarzschild time as $a\rightarrow0$ and the spacetime
approaches the Schwarzschild spacetime. It follows that the Bowen-York
method of generating initial value solutions, when applied to configurations
with initial Kerr holes, or for the close-limit approximation method, entails
the difficulties outlined in Sec.\,I.

It is natural to ask, aside from the ``practical'' questions related
to black hole collisions, whether any of the restrictions in our
conclusion can be modified or removed, so that a more general
conclusion can be stated about conformally flat slicings of the Kerr
spacetime.  The axisymmetric restriction might appear to be
particularly simple to remove, since we use what amounts to a
perturbation (in $a$) expansion in which the ``background'' (the Kerr
spacetime) is axisymmetric.  But some of the terms in the Bach tensor
are second order in the slicing funtion $F$, so a Fourier decomposition of
$F(r,\theta,\phi)$ will result in a mixing of Fourier modes.  A
solution without a Fourier decomposition would appear to be quite
difficult. The key to the relatively simple result we have presented
is that the diagonal components of the Bach tensor, to second order in
$a$, are linear in $F$ for axisymmetric slicings. That simplicity
disappears in we allow for $\phi$ dependence in $F$, and greatly
complicates an approach of the type we have used. Since this is a
difficult task and has little connection with questions of initial
data sets, we have not pursued it.

An attempt to look at slicings with the form in eq.\,(\ref{gensliceq})
runs into different problems. In this case, for nontrivial $G$, the
Bach tensor will have terms to first order in $a$. These terms will be
linear in both $G$ and in $F$.  Choices of $G$ and $F$ for which the
slicing is conformally flat to first order in $a$ cannot be ruled
out. (If, for example, one chooses both $F$ and $G$ to be functions
only of $r$, the slicing to first order in $a$ is spherically
symmetric and hence conformally flat). It follows that from the first
order Bach terms one can only infer
 restrictions on $G$ and on $F$.
These restrictions must be applied to the equations that 
arise from the terms to second order in $a$, equations that include 
terms quadratic in $F$. Again, we have not pursued a generalization 
along these lines.

\acknowledgments We thank Walter Landry, William Krivan and Karel
Kucha\v{r} for useful discussions. We thank James Bardeen for bringing 
to our attention the possibility that the Schwarzschild geometry
may have slices that are conformally flat other than those of constant
Schwarzschild time.
Special thanks go to John Whelan
for helpful suggestions at the early stages of this work. We
gratefully acknowledge the support of the National Science Foundation
under grant PHY9734871.

\end{document}